\documentclass[]{spie}  

\usepackage{amsmath,amsfonts,amssymb}
\usepackage{graphicx}
\usepackage[colorlinks=true, allcolors=blue]{hyperref}

\usepackage{gensymb}

\title{Near UV Imager with an MCP Based Photon Counting Detector}

\author{S.~Ambily\supit{a}, Joice Mathew\supit{a}, Mayuresh Sarpotdar\supit{a},  
A.~G.~Sreejith\supit{a}, K.~Nirmal\supit{a}, Ajin Prakash\supit{a}, Margarita Safonova\supit{a} and
Jayant Murthy\supit{a}
\skiplinehalf
\supit{a}Indian Institute of Astrophysics, Koramangala 2nd Block, Bangalore,
India
}

\authorinfo{Send Correspondance to: ambily@iiap.res.in}

\begin{document} 
\maketitle

\begin{abstract}

We are developing a compact UV Imager using light weight components, that can be flown on a small CubeSat or a balloon platform.  The system has a lens-based optics that can provide an aberration-free image over a wide field of view. The backend instrument is a photon counting detector with off-the-shelf MCP, CMOS sensor and electronics. We are using a Z-stack MCP with a compact high voltage power supply and a phosphor screen anode, which is read out by a CMOS sensor and the associated electronics. The instrument can be used to observe solar system objects and detect bright transients from the upper atmosphere with the help of CubeSats or high altitude balloons. We have designed the imager to be capable of working in direct frame transfer mode as well in the photon-counting mode for single photon event detection. The identification and centroiding of each photon event are done using an FPGA-based data acquisition and real-time processing system. 
\end{abstract}


\section{INTRODUCTION}
\label{sec:intro} 

The advent of better technologies capable of putting a payload in the outer layers of the Earth's atmosphere has made it relatively easier to develop and fly compact scientific instruments for astronomical and atmospheric studies. CubeSats and high altitude balloons provide a good opportunity to make significant observations at the relatively under-explored regions of the UV spectrum\cite{Noah}. We are currently in the process of developing such payloads that work in the near UV region, using commercially available off-the-shelf components and with tight constraints on weight, size, and power. The high altitude ballooning program at the Indian Institute of Astrophysics can be used to reach up to heights of 30km for initial testing of the payload components\cite{Safonova}. We are also planning to use the National Balloon Facility station of TIFR at Hyderabad which can float at 40 km for up to 5 hours using zero-pressure balloons\cite{tifr} 

One of the current payloads that we have in the design and development stage is a wide field imager for the near UV region. We are using a lens system made of CaF\textsubscript{2} and Fused Silica so that there's good transmission above 250nm. Since the photocathode of the MCP has good a spectral response from 200 to 900nm, we can use it in both NUV and visible regions with suitable filters. The detector can work in photon counting mode for observing faint UV sources from the upper atmosphere. Under good light levels, the detector works as an intensified CMOS camera, transferring each frame serially to the ground station or storing them on the payload for later analysis. In both the cases, the data acquisition and processing card is the same FPGA board, which implies that the modes can be changed by simply switching from one program to the other in the FPGA, without making any changes to the hardware.

In the following sections, we describe the various aspects of the design and implementation of the system from the design of optics, choice of components, an overview of the design flow and algorithm. 

\section{Design}
We are planning to put the instrument on a CubeSat platform after the initial testing and so, the design of the optical and electronics assembly is being done keeping in mind that the payload should fit in a 200mm x 200mm x 100mm box. 

\subsection{Optics}
The imager is intended to work at a wavelength range of 250-400 nm. We have designed a modified double-Gauss lens system with CaF\textsubscript{2} and Fused Silica lenses for focussing the light to the MCP. The spot size is adjusted such that the PSF is spread over 4 pores on the MCP. The lens design parameters are summarized in Table ~\ref{table:lens}

\begin{table}[ht!]
\centering
\caption{Lens specifications} 
\begin{tabular}{ll} 
\hline
\rule[-1ex]{0pt}{3.5ex} Number of Elements & 5 \\
\rule[-1ex]{0pt}{3.5ex} System Aperture & 70 \\
\rule[-1ex]{0pt}{3.5ex} Total Track & 250.39 \\
\rule[-1ex]{0pt}{3.5ex} Working F Number  &  2.98 \\
\rule[-1ex]{0pt}{3.5ex} Paraxial Image Height & 19.76 \\
\rule[-1ex]{0pt}{3.5ex} Maximum Radial Field &  5.4 \\
\rule[-1ex]{0pt}{3.5ex} Materials & CaF\textsubscript{2} and Fused Silica \\
\hline
\end{tabular}
\label{table:lens}
\end{table}

\begin{figure}[ht]
\centering
\includegraphics[width=0.5\textwidth, angle =90]{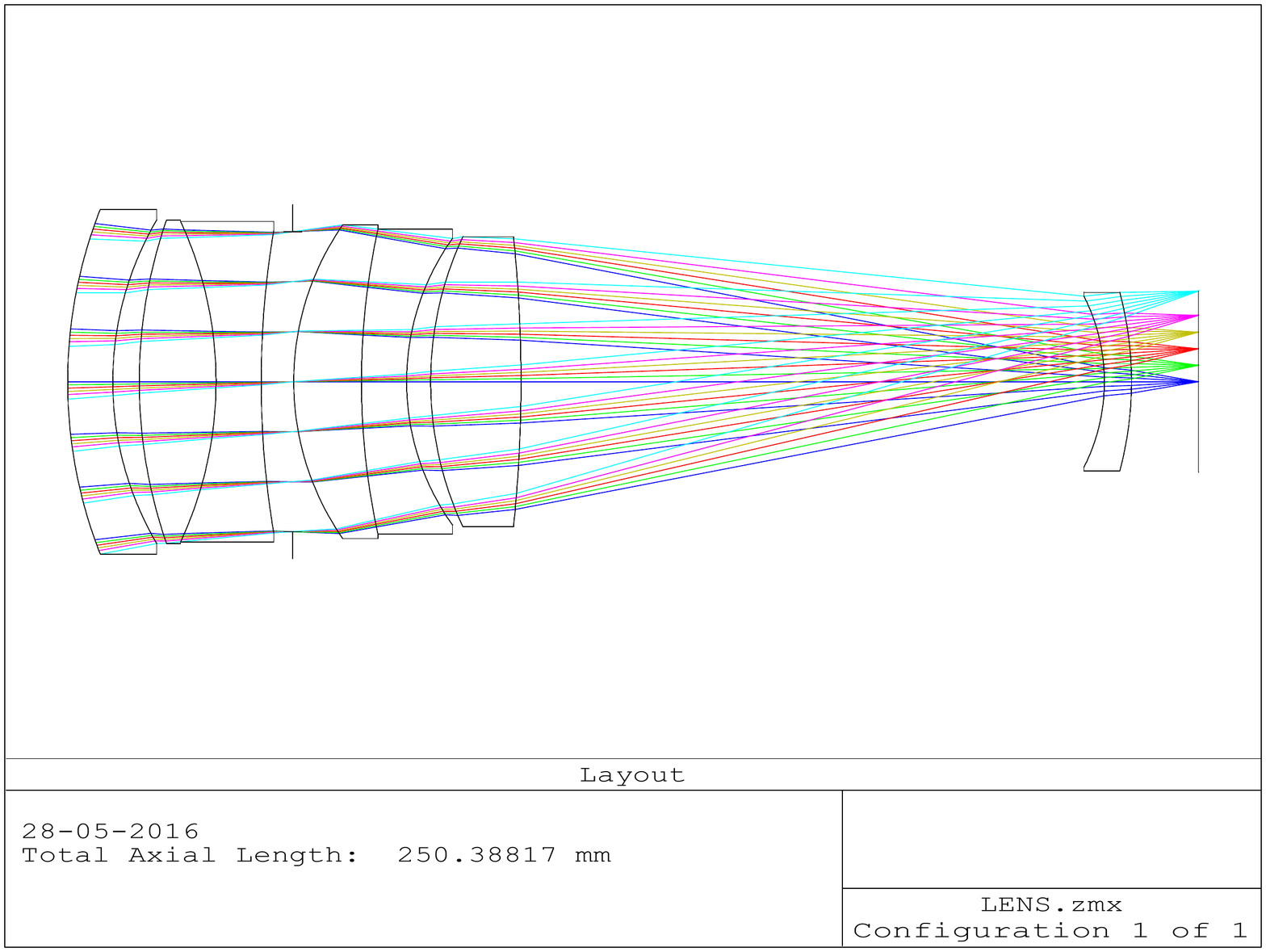}
\includegraphics[width=0.5\textwidth, angle =90]{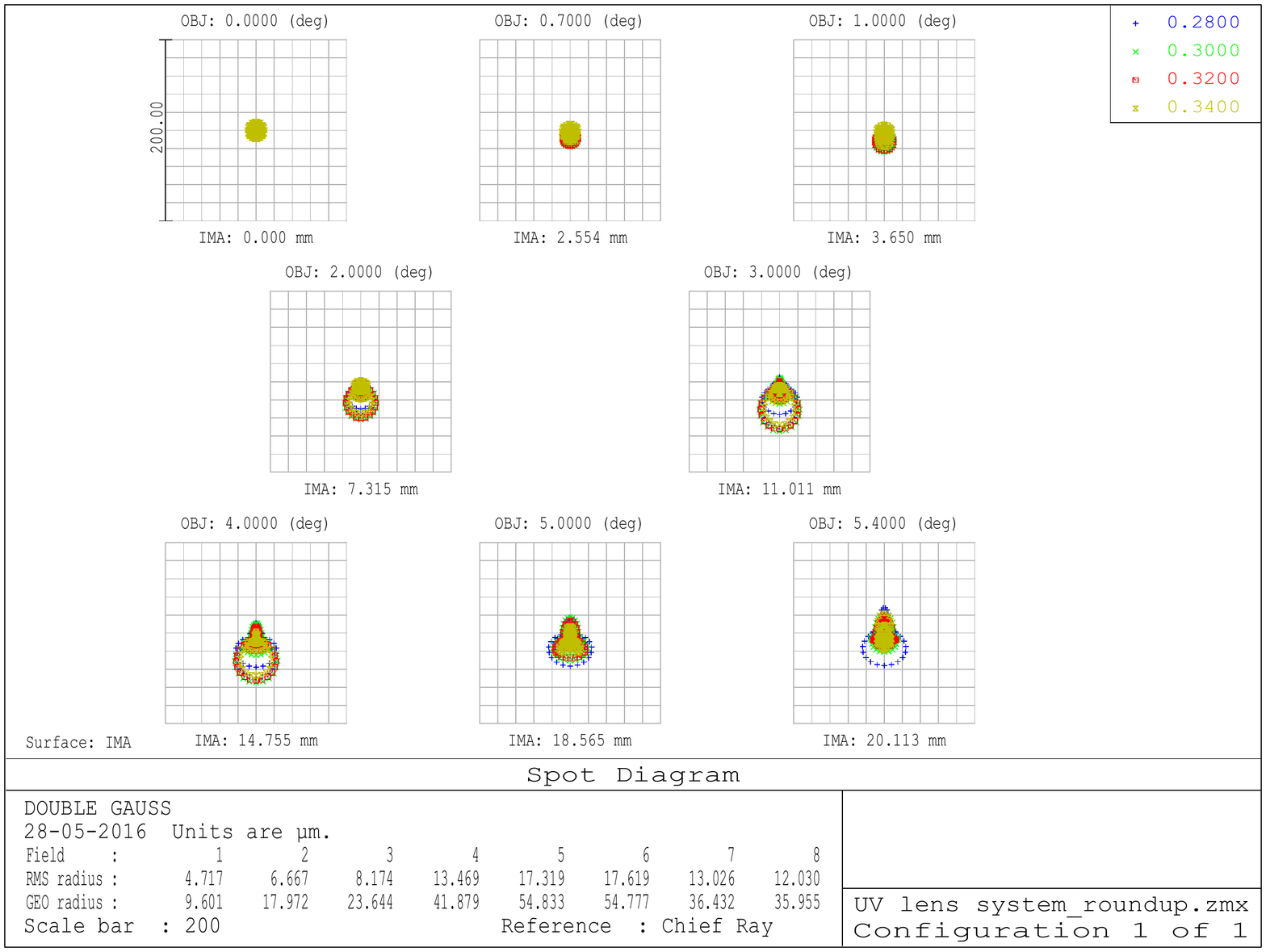}
\caption{{\it Top}: Layout of the optics; {\it Bottom}: Spot Diagrams for a 10.8\degree field.} 
\label{fig:result1}
\end{figure}

As shown in the table, currently the length of the track exceeds our limit of 200mm. We are solving this by placing 2 fold mirrors before the image plane so that the track is folded back to fit in the allotted space. 

\subsection{Detector}
Photon-counting detectors using MCPs have been the standard for UV payloads because of their low readout noise, large detector area, radiation tolerance and long-wavelength rejection\cite{Kimble} \cite{Vallerga_general}. The detector parameters such as sensitivity, gain, spatial resolution depends on the variations in the number of the MCP stacks, types of filters and photocathodes, and in the anode and readout electronics\cite{Joseph}. All-electronic readout systems such as delay line anode, cross strip anode, .etc. can be read out by custom-developed Application Specific Integrated Circuits(ASIC) to realize higher readout rates and lesser distortion\cite{Siegmund_galex}. Since the higher cost and lead times for implementing an ASIC based readout for the MCP is prohibitive for our purposes, we are implementing an optical readout of the MCP with a focussing lens system, a CMOS sensor and the electronics board\cite{Uslenghi_2}. 

\begin{figure}[ht]
\centering
\includegraphics[scale=0.6]{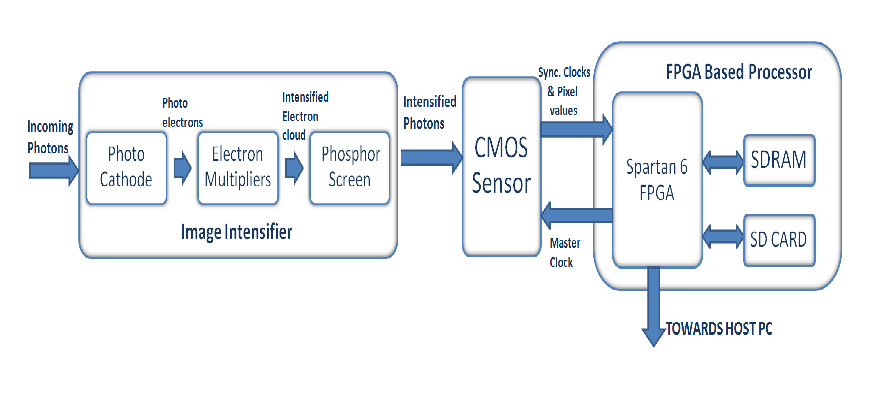}
\caption{Block Diagram of the Detector Setup}
\label{fig:blockdia}
\end{figure}

\subsubsection{MCP Assembly}
We are using the MCP340 assembly from Photek\footnote{\tt{http://www.photek.com/}}, which is a 40-mm diameter 3-stage MCP. It includes a quartz input window, S20 photocathode, Z-stack MCP and a P46 phosphor screen. The MCP requires a high-voltage power supply (HVPS) to operate in the single-photon mode, for which we have used the FP632 micro-HVPS from Photek. The power supply weighs only 100 gm, which makes it attractive for our lightweight balloons, where the typical payload weight is less than 2 kg. The MCP voltage levels are kept at $-200, 0, 2300$ and $6600$ V for the cathode, MCP-In, MCP-Out and phosphor screen, respectively. The gain of the MCP is set by the voltage at the MCP output, which automatically adjusts the screen voltage, maintaining a constant MCP-out-Anode voltage difference. The resulting cloud of electrons from the MCP output electrode is accelerated towards the P46 phosphor screen anode (with a peak response at 530 nm), and the photons from the anode are focused on the CMOS surface through a relay lens. 

\subsubsection{Focusing Optics}
This involves coupling the light from the MCP to the CMOS chip through a lens system, or a fiber optics taper, with minimum distortion and light loss. Since the MCP is 40mm in diameter and the CMOS sensor is 18mm x 14mm, we need a demagnification of 3:1 from the optics. In addition, the tube length of the entire detector assembly is to be confined to 150mm. We are designing a 4 element lens system with tight constraints on the focal length. We have tried to keep the distortion introduced by the lens to a minimum, but further analysis and modeling of the distortion is required for correcting these effects through in real-time.  

\subsubsection{CMOS Sensor}
We are planning to use the NOIL2SM1300A which is a 1.3 MP sensor from On Semiconductors\footnote{\tt{http://www.onsemi.com/}}. It's a high-sensitivity, high-speed sensor with capabilities for sub-sampled, windowed outputs, which helps us improve the event counting rate of the detector. We have developed a generic FPGA-based readout board\cite{mico} for CMOS sensors which can be used as the master control board for the sensor. The board takes care of functions such as providing the bias and clock for the sensor, programming of the FPGA, and data storage and transmission. The use of FPGAs allows us to rapidly prototype different algorithms without changing the hardware which is invaluable in a development environment. The main processor of the board is realized on a military grade Spartan\textsuperscript{\textregistered} 6Q series XQ6SLX150 chip\footnote{\tt{Xilinx, Inc: http://www.xilinx.com/}}, that takes care of all the basic functions like clocking and readout of the CMOS, centroiding of events, interfacing the on-board SDRAM and SD card for storage of data and telemetry and communication with the control interface.  

Technical specifications of detector components are summarized in Tables ~\ref{table:detector}

\begin{table}[ht!]
\centering
\caption{Detector specifications} 
\begin{tabular}{ll} 
\hline
\rule[-1ex]{0pt}{3.5ex} MCP Diameter & 40 mm \\ 
\rule[-1ex]{0pt}{3.5ex} Type of MCP & Z stack \\
\rule[-1ex]{0pt}{3.5ex} Photocathode & S20 \\
\rule[-1ex]{0pt}{3.5ex} Anode Type & P46 Phosphor screen \\
\rule[-1ex]{0pt}{3.5ex} Pore size & $10\,\mu$m (diameter) and $12\,\mu$m (pitch)\\
\rule[-1ex]{0pt}{3.5ex} Input Window & Fused Silica \\
\rule[-1ex]{0pt}{3.5ex} Output Windows & Glass \\
\rule[-1ex]{0pt}{3.5ex} Cathode Output Voltage (Max) & $-220$ V \\
\rule[-1ex]{0pt}{3.5ex} MCP Output Voltage (Max) & 2800 V \\
\rule[-1ex]{0pt}{3.5ex} Anode Output Voltage (Max) & 6000 V \\
\rule[-1ex]{0pt}{3.5ex} CMOS Active Array Size & $1280\times 1024$ (H$\times$V) \\
\rule[-1ex]{0pt}{3.5ex} Frame Rate & 500 fps @$1280\times 1024$ \\
\rule[-1ex]{0pt}{3.5ex} CMOS Master Clock & 315 MHz \\
\rule[-1ex]{0pt}{3.5ex} Data Format & 10 bit parallel (RGB) \\
\rule[-1ex]{0pt}{3.5ex} Pixel Size & $14\,\mu$m $\times 14\,\mu$m \\
\rule[-1ex]{0pt}{3.5ex} Dynamic Range & upto 90 dB \\
\rule[-1ex]{0pt}{3.5ex} CMOS Operating Temperature & -50°C to +85°C \\
\rule[-1ex]{0pt}{3.5ex} Type of FPGA & Spartan 6Q (XQ6SLX150T) \\
\rule[-1ex]{0pt}{3.5ex} Number of Logic Cells &  147000 \\
\rule[-1ex]{0pt}{3.5ex} On-board SDRAM & 512 MB \\
\rule[-1ex]{0pt}{3.5ex} FPGA Temperature Range & -40 to 100°C  \\
\hline
\end{tabular}
\label{table:detector}
\end{table}

\section{Implementation of prototype detector}
We are currently in the process of developing and testing a prototype detector for the imager. This is useful for calibrating the MCP assembly and the FPGA algorithms. We have also developed the data acquisition and centroiding algorithms on the FPGA to be tested with the system. The prototype detector consists of off-the-shelf boards for the CMOS sensor and FPGA. We have chosen the OV9715, a 1 mega pixel video image sensor from OmniVision Technologies\footnote{\tt{http://www.ovt.com/}}, for the CMOS camera. The CMOS chip sits on an off-the-shelf headboard, originally designed for use in CCTV applications\footnote{\tt{Soliton Technologies: http://www.solitontech.com/}}. All the electronic components for biasing and clocking of the CMOS sensor are present on-board. The headboard reads the digital image values and sends them to the FPGA for further processing. 

We are using the XuLA2-LX25 FPGA prototyping board from XESS Corp.\footnote{\tt{http://www.xess.com/shop/product/xula2-lx25/}} for the design of the main processor. This board houses a Spartan 6 XC6SLX25 FPGA and has the necessary components such as SDRAM(synchronous dynamic random-access memory), SD Card and clocking resources in a compact form factor to control and read out the CMOS sensor, process the data, and store and transmit the output. The FPGA modules were written using Verilog and synthesized and programmed using the Xilinx ISE\textsuperscript{\textregistered} (Integrated Synthesis Environment) Design Suite. In addition, we have used standard VHDL modules\footnote{\tt{XESS Corp.}} for interfacing the SDRAM, PLL(Phase Locked Loop), and micro-SD(Secure Digital) card to the Spartan\textsuperscript{\textregistered} 6 chip on the FPGA board. 

\section{Centroiding}
The Gaussian charge cloud from the MCP is spread over multiple pixels on the CMOS, and our aim is to find the exact centroid of the event with sub-pixel accuracy. Centroiding improves the spatial accuracy with which the photon hits are identified and helps reduce the data bandwidth requirements. The key challenge here is to get the most accurate method of centroid computation while using the limited resources of the FPGA. It is not practical to use the highly complicated development boards from Xilinx due to the tight constraints on the weight and dimensions of the payloads.The centroiding was implemented using a $3\times 3$ pixel sliding window at first for simplicity, and later were expanded to $5\times 5$ pixels window, since they were found to yield better accuracy with the minimum amount of correction\cite{Hutchings}. 

\begin{figure}[ht!]
\begin{center}
\includegraphics[scale=0.6]{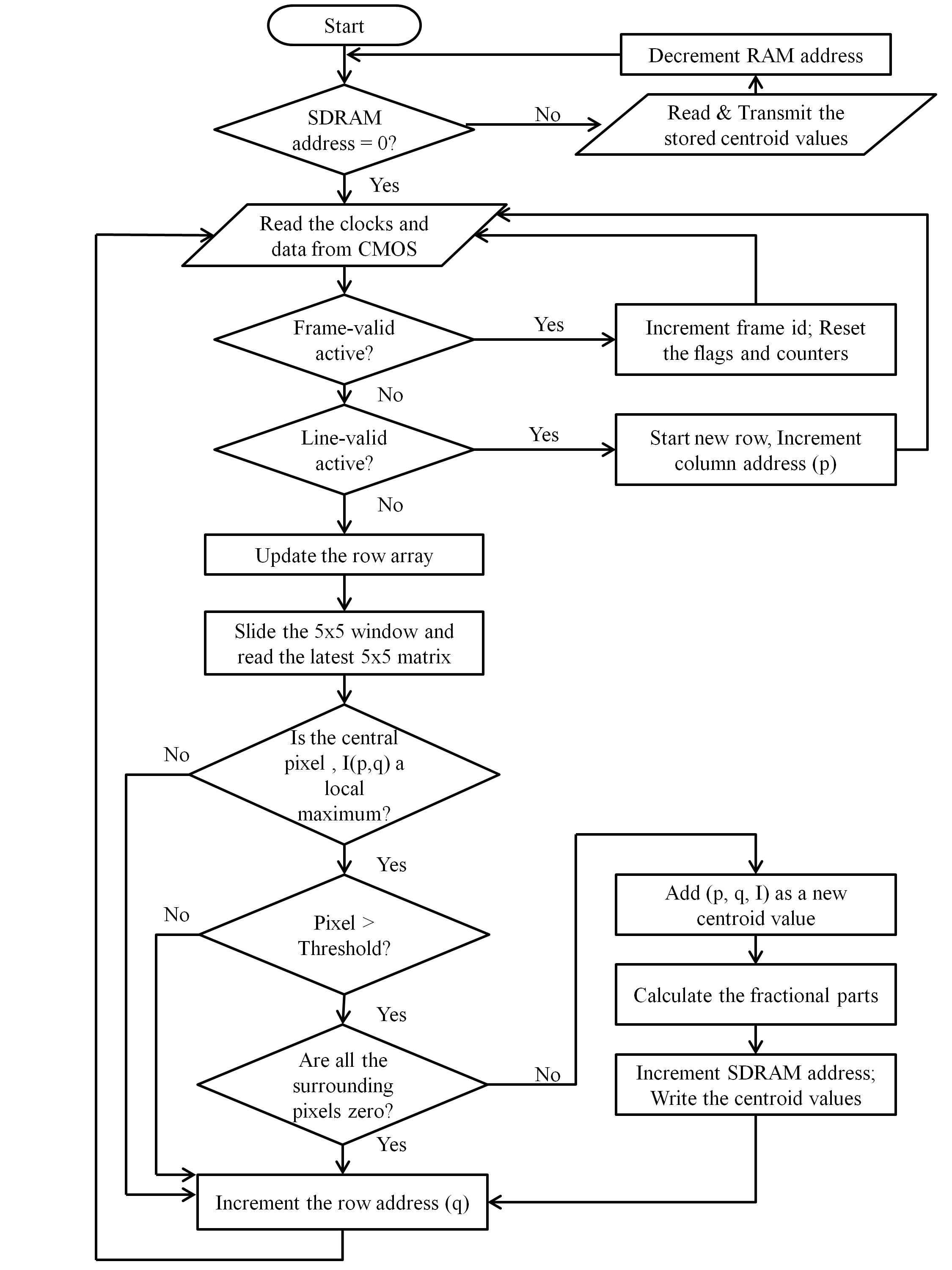} 
\end{center}
\caption{Centroiding Flowchart}
\label{fig:flowchart}
\end{figure}

The limited number of logic cells on the FPGA board means we need to optimize the algorithm without compromising on the speed. Even with a reduced image size of $640 \times 400$, it is not possible to store all the pixel values of the image on the FPGA chip itself. In the continuous frame transfer mode, all the pixel values even for the $1280 \times 800$ image can be stored on the SDRAM module on the board. But while centroiding, the time taken to access the RAM for each byte was seen to contribute to a significant delay in the processing. This was solved by creating a smaller array of 5 rows on the FPGA chip ($5 \times 1280$), where only the most recent five rows of the current frame are stored. 

As the first step of centroiding, we need to apply a threshold that is obtained from the average background noise level in the image. This can be a programmable number or computed dynamically from each of the incoming frames. We have done the simulations to test this algorithm on MATLAB\textsuperscript{\textregistered} and found that the optimum threshold is equal to, or slightly higher than, the mean signal level of the image for detecting the maximum number of photon events with the best output SNR. As and when each pixel was read out, the array containing the last five rows was updated, while the older row values are overwritten. The 24 pixels surrounding the current one (say, $p-2, q-2$) are read into a separate 5$\times$5 window array, and the window is checked for the presence of any local maxima. We had to flag hot pixels and multiple events before calculating the centroids as well. Once we have identified the hot pixels and multiple events, we calculate the centroids for the actual photon events in the current row. We saved the coordinates of the central pixel as the integer parts of the centroid and computed the sub-pixel values in parallel by calculating the difference in intensities between the neighboring top and bottom pixels and the left and right pixels of the event. 

\begin{figure}[ht]
\centering
\includegraphics[width=0.999\textwidth]{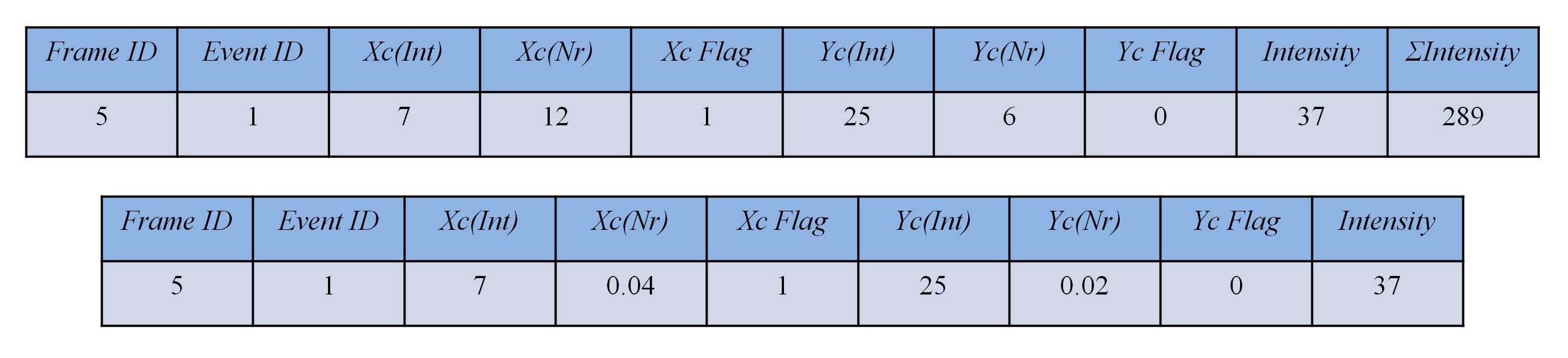}
\caption{Data packet for each photon event. The final output packet is optimized to contain 7 or 8 bytes depending upon the sub-pixel accuracy required.}
\label{fig:row3}
\end{figure}

In the centroiding mode, the output data for telemetry is in the form of packets. Each packet contains a frame ID, event ID, integer and fractional values of centroid coordinates, and the intensity value of the central pixel (Fig.~\ref{fig:row3}). These packets are saved on the FPGA by creating a block of registers on the chip which results in a faster operation. However, the number of photon events that can be saved per frame is limited by the number of logic cells available on-chip. To solve this, we save each packet to the SDRAM as and when it is generated.

Since it is impossible to realize a divider module in FPGA hardware, Hardware description languages do not support division operation and division operations are accomplished using repeated subtractions which add cost to the time and resources. Therefore, when transmitting the packet from the centroiding to the telemetry module, the sub-pixel values are not computed but are simply sent as the numerator and the denominator values. Before transmission, the sub-pixel values can be calculated with an accuracy of 4 bits each by a separate divider module which can be increased to 6 or 8 bits at the expense of extra clock cycles. 

The centroiding algorithm is explained in the flowchart (Fig.~\ref{fig:flowchart}). 

\section{Summary and Future Work}
We have developed a basic electronic readout system for the CMOS chip using FPGAs along with the data acquisition and centroiding algorithms. The performance of the system is predominantly limited by the performance of the FPGA chip and the speed of CMOS data transfer as of now. We are doing the testing and calibration of the MCP and the power supply, with the present readout mechanism. We also are in the process of procuring components with higher performance for the final version of the detector and evaluating the same algorithms on them. 

Another important work ahead is to make the whole detector system flight-ready. We have to complete the space-qualified optical and mechanical assembly and do the tests. In addition, the performance of all the mil-grade electronic components for radiation-prone environments needs to be checked. The distortion effects from the focussing lens need to be studied and modeled so that corrections can be applied to the output images. When operating the MCP in the high-altitude balloon flights, a better temperature control needs to be maintained. 

\section{Acknowledgments}
Part of this research has been supported by the Grant IR/S2/PU-006/2012 from the Department of Science and Technology, Government of India.

\bibliography{master_file}{}
\bibliographystyle{spiebib}
\end{document}